\documentstyle[pra,aps,multicol,epsfig,amssymb]{revtex}

\title{Non--Interacting Fermions in a One--Dimensional Harmonic
Atom Trap: Exact One--Particle Properties at Zero Temperature}

\author{F. Gleisberg and W. Wonneberger} 

\address{Abteilung f\"ur Mathematische Physik, Universit\"at Ulm, D89069 Ulm, Germany}

\author{U.~Schl\"oder and C.~Zimmermann }

\address{Physikalisches Institut, Eberhard Karls Universit\"at, D72076 T\"ubingen, Germany}

\date{May 19, 2000}

\begin{document}
\maketitle

\begin{abstract}
One--particle properties of non--interacting Fermions in a one--di\-men\-sio\-nal 
harmonic trap and at zero temperature are studied.
Exact expressions and asymptotic results for large Fermion
number $N$ are
given for the particle density distribution $n_0(z,N)$. For large
$N$ and near the classical boundary at the Fermi energy the density displays 
increasing fluctuations. A simple scaling of these tails of the density distribution 
with respect to 
$N$ is established. The Fourier transform of the density distribution is
calculated exactly. It displays a small but characteristic hump near 
$2 k_F$ with $k_F$ being a properly defined Fermi wave number. This is due
to Friedel oscillations which are identified and discussed. These quantum
effects are missing in the semi--classical approximation.  
Momentum distributions are also evaluated and discussed. As an example of 
a time--dependent one--particle problem we calculate exactly the evolution 
of the particle density when the trap is suddenly switched off and 
find a simple scaling behaviour in agreement with recent general results. 
\end{abstract}

\vspace{0.5cm}

\hspace*{2.2cm}PACS  Nos. 05.30FK, 71.10.Pm, 0375.Fi
\begin{multicols}{2}
\narrowtext

\section{Introduction}

Recent years brought about spectacular successes in the study of 
dilute bosonic quantum gases confined 
to atomic traps at extremely low temperatures. These and the 
experimental details are reviewed in \cite{KDSK99}.

The next stage of investigations will incorporate fermionic 
quantum gases. Fermi degeneracy of potassium atoms (${\rm^{40}K}$) has 
recently been 
observed in \cite{DMJ99}. 
The effects of interactions between the neutral atoms are of 
particular interest. They
can give rise to collective ground states like superfluid phases.

Another development
regards the construction of highly anisotropic traps, e.g., the 
microtraps in \cite{VFP98,FGZ98,DCS99,TOZ99,RHH99}. The magnetic
trapping fields can be taylored so as to make the confining potential
harmonic.

If the longitudinal
confinement frequency $\omega _{\ell}$ is smaller than the 
radial frequency $\omega _r$ by a factor $\lambda$
it is possible to fill the first 
$ N$ longitudinal states while the radial 
 wave functions of the Fermions are still that of the ground state
provided $N < 1/\lambda$ holds.

From semi--classical theory or local density approximation (LDA) it is well known 
(cf. e.g. \cite {BR97}), that a Fermi wave number $k_F=\sqrt{(2N-1)m \omega _{\ell}/\hbar}$
can be associated with the one--dimensional Fermi gas of atomic mass $m$
in a harmonic trap. It is noted that the condition $N < 1/\lambda$
is roughly in line with the standard estimate $k_F < 1/l_t$ for a Fermi 
system which is confined to a transverse width $l_t$ to be quasi one--dimensional 
provided the length $l_t$ is identified with the extension $l_r=\sqrt
{\hbar/(m \omega _r)}$ of the radial ground state wave function in the trap. 

Non--interacting Fermions in anisotropic harmonic traps have been studied 
recently \cite {BR97,SW98,BC00} using exact and semi--classical methods. The 
thermodynamics of harmonically confined spin--polarized Fermions in any spatial dimension
including a harmonic two--particle interaction has been studied in \cite{BDL98}
using the general approach \cite{BDL97}. The latter results are not available in
closed form and require numerical evaluation. A finite series re\-pre\-sen\-tation 
for the free energy of one--dimensional non--interacting spin--polarized Fermions 
confined by a harmonic potential has been given in \cite{TI83}. 

In view of the feasibility to realize one--dimensional Fermions at ultra--low temperatures 
it seems worthwhile to supplement these works by studying the strictly one--dimensional 
case of non--interacting Fermions at zero temperature when a number of exact explicit 
results can be obtained.
 
Interactions between spin--polarized identical Fermions are weak because 
the Pauli principle forbids s--wave scattering. On the other hand the theory of 
Luttinger liquids (cf. e.g., \cite{V95} for a review) shows that even small interactions 
change a one--dimensional Fermi system substantially. Nevertheless, it is useful 
to have results for the non--interacting case to compare the effect of interactions 
with them. The results that we present below show features specific for one spatial 
dimension.

In existing micro traps magnetic gradients of up to $30$ T/cm has already
been realized \cite{VFP98} resulting in a periodic motion of the trapped atoms on
a time scale of micro seconds.  Novel versions of micro traps based on
micro fabricated current conductors achieve even higher gradients with
an expected radial atomic oscillation frequency of above $1$ MHz \cite{TOZ99}.  For
the longitudinal oscillation frequency $1$ Hz appears to be a reasonable
lower limit because time scales longer than a second gives rise to
experimental difficulties due to seismic and acoustic noise.  Thus, the
maximum value which is currently feasible for $\lambda$ is $10^{-6}$ and would limit
the number of atoms inside the trap to about one million.  The main
experimental difficulty, however, is to fill the $10^6$ states of the micro
trap with a substantial number of atoms.  Starting from an optically
cooled sample of atoms with a phase space density of typically $10^{-6}$ 
\cite{TEC95} a phase space compression of six orders of magnitude is required
to completely fill up the wave guide.  Such compression is possible with
state of the art techniques of evaporative cooling \cite{LRW96}.  Thus, a
conservative estimation for realistic experimental conditions would
assume a one component fully spin polarized Fermi--gas with a radial
frequency inside the micro trap of $10^5$ Hz. 
The longitudinal frequency can be set at $10$ Hz 
giving $\lambda =10^{-4}$. Thus $N=10^4$ quasi one--dimensional Fermions can be
accommodated inside the trap. Assuming ${\rm ^6Li}$ atoms (in the hyperfine 
state $|m_s=1/2, m_i=1 \rangle$) the inverse harmonic oscillator length $\alpha$ 
according to

\begin{equation}\label{0.1}
\alpha = \sqrt{m \omega _{\ell} /\hbar}
\end{equation}

is estimated as $\alpha \approx 8 \cdot 10^2 \mbox{ cm}^{-1}$ leading to a Fermi wave
number $k_F \approx 10^5 \mbox{ cm}^{-1}$. 

Obviously the quasi one--dimensional Fermi energy $\epsilon _F$, i.e., the
energy of the highest occupied state without the radial contribution is

\begin{equation}\label{0.2}
\epsilon _F =\hbar \omega _{\ell} (N- \frac{1}{2}).
\end{equation}

Under the above assumptions $\epsilon _F$ corresponds to about 
$5\, \mu$K and this 
temperature must be larger than the physical temperature in
order to achieve degeneracy of the Fermi gas.

Another relevant quantity is the spatial extension of the inhomogeneous
Fermi gas. The appropriate measure is twice that later given in equation 
(\ref{2.1})
and leads to a characteristic extension of $0.4$ cm and to
an average Fermion density of about $3 \cdot 10^4$ atoms per cm. The radial
width $2 l_r$ is about $3 \cdot 10^{-5}$ cm. Thus the tonks gas limit 
\cite{O98} is avoided and the fermionic atoms can be treated as 
point particles.

The exact quantum mechanical results usually give only small corrections
to the corresponding LDA predictions. Some of them are, however, of 
qualitative nature and worth to point out. Among them are 
diverging density oscillations near the classical boundary
of the trap for large Fermion numbers and the general feature of Friedel
oscillations \cite{JF54} of the density. 

The paper is organized as follows. Sec. II
presents the basic theory. Sec. III discusses the relevant lengths and energy scales
of the one--dimensional Fermi gas in the harmonic trap. In Sec. IV we compile the  
results for the zero temperature one--particle density distribution.
Sec. V is concerned with the Fourier transform of the
density distribution. Sec. VI discusses momentum distributions and in the 
final Sec. VII we calculate the expansion of the particle density distribution
when the trap is suddenly switched off. An Appendix
summarizes mathematical formulae used in our calculations.

\section{Basic Theory}

We consider a gas of spinless non--interacting Fermions in one 
spatial dimension and trapped in a harmonic potential

\begin{equation}\label{1.1}
V(z) = \frac{1}{2} m \omega ^2 _{\ell} z^2.
\end{equation}

The Hamiltonian in second quantization and for the grand 
canonical ensemble is

\begin{equation}\label{1.2}
\hat{H}_0 = \sum^\infty _{n=0} (\hbar \omega _n
- \mu) \hat{c}^+_n \hat{c}_n
\end{equation}

 with one--particle energies $\hbar \omega _n = \hbar \omega _{\ell} (n+1/2), n=0,1,...$.
 The chemical potential is denoted $\mu$. The Fermion creation operators 
 $\hat{c}^+$ and destruction operators $\hat{c} $ obey the fermionic algebra
 $ \hat{c}_m \hat{c}_n^+ + \hat{c}_n^+ \hat{c}_m = \delta _{m,n}$.   
 This ensures that each (non-degenerate) energy level $\epsilon _n = \hbar \omega _n $ with 
 (real) single particle wave function
 
 \begin{equation}\label{1.3}
 \psi _n(z) = \sqrt{\frac{\alpha}{2^nn!\pi^{1/2}}} 
 \,e^{-\alpha^2 z^2/2 }\,H_n (\alpha z)
 \end{equation} 

 (normalized according to $ \langle m |n \rangle = \delta _{m,n}$) 
 is at most singly occupied. The 
 intrinsic length scale of the system is the oscillator length $l = \alpha ^{-1}$ 
 where $\alpha$ is defined by (\ref{0.1}). $ H_n$ denotes a Hermite polynomial.

 We consider the spatial density of one--dimensional Fermions in the harmonic trap, i.e.,
 the one--particle distribution function

 \begin{equation}\label{1.5}
 n(z;T,\mu) = \mbox{Tr}\, \hat{\rho} \,\hat{\psi}^+ (z) \hat{\psi} (z).
 \end{equation}
 
 In (\ref{1.5}) the operator $\hat{\psi}(z) $ destroys a Fermion at position $z$. It can 
 be expanded as
 
 \begin{equation}\label{1.6}
 \hat{\psi} (z) = \sum^\infty _{n=0} \psi _n (z)\, \hat{c}_n.
 \end{equation}
 
 The density operator is
 
 \begin{equation}\label{1.7}
 \hat{\rho} = Z^{-1} e^{-\beta\hat{H}_0}
 \end{equation}
 
 with $Z =\mbox{Tr} \,\exp[-\beta \hat{H}_0]$. 
 A standard textbook exercise then gives
 
 \begin{eqnarray}\label{1.9}
 n_0 (z;T,\mu) = \sum^\infty _{m=0} \psi ^2_m (z)\,\,p_m (T,\mu) 
 \end{eqnarray} 

 where
 
 \begin{eqnarray}\label{1.10}
 p_m(T,\mu) =  \left \{ e^{ \beta(\hbar \omega _m -\mu)} 
 + 1 \right \}^{-1}
 \end{eqnarray}

 is the thermal occupation number of the single particle state $\psi _m$.
  
 The present paper deals with the case $T \rightarrow 0 $ when a number of 
 analytical results are available. The important simplification results 
 from the fact that for $T \rightarrow 0$ the first $N$ levels are completely 
 filled while all others are empty, i.e.,

 \begin{equation}\label{1.11}
 p_m(T \rightarrow 0,\mu)  \rightarrow \Theta(N-1-m)
 \end{equation}
 
 and $\mu$ becomes the Fermi energy $\epsilon _F$:
 
 \begin{equation}\label{1.12}
 \mu \rightarrow \epsilon _F = \hbar \omega _{\ell} (N-\frac{1}{2}).
 \end{equation}
 
 The density $n_0 (z; T \rightarrow 0,\mu)$ which we henceforth denote $n_0 (z, N)$ takes 
 on the form
 
 \begin{equation}\label{1.13}
 n_0 (z,N) = \sum^{N-1}_{n=0} \psi _n^2 (z).
 \end{equation}

 The zero temperature case is depicted in Fig. 1.

 \begin{figure}[h]
\begin{center}
 \epsfig{figure=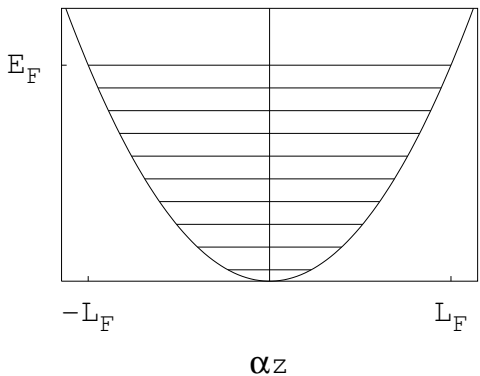,width=0.9\columnwidth} 
\end{center}
 \caption{\small$N=10$ non--interacting spinless Fermions filling the lowest single particle 
 levels in the harmonic trap at zero temperature. $\epsilon _F$ denotes the Fermi energy and $L_F$ the half
 width of the Fermi system.}
 \end{figure}

 Equation (\ref{1.13}) is the main object of the present study. Under the condition
 $k_B T \ll \epsilon _F$ it correctly describes the density of non--interacting 
 Fermions in a harmonic trap.

 Fig. 2 shows the density profile with the characteristic ripples on top.

 This is in contrast to an
 infinite Fermi gas (or one with periodic boundary conditions) where the density 
 is homogeneous. The ripples appear here as a finite size effect. In the 
 center of the trap they will be identified below as Friedel oscillations \cite{JF54}.

 \begin{figure}[ht]
\begin{center}
\epsfig{figure=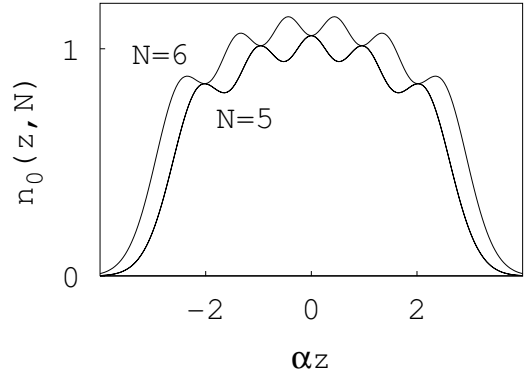,width=0.9\columnwidth} 
\end{center}
 \caption{\small Particle density distribution functions in units of the inverse oscillator length
 $\alpha$ for $N=5$ and $N=6$ Fermions in a 
 one--dimensional harmonic trap and at zero temperature. The added 
 Fermion resides in the area between the two curves. The density oscillations near 
 the center can be identified as Friedel oscillations.}
 \end{figure}

\section{Lengths and Energy Scales}

 In this Section we summarize the relevant scales of the one--dimensional
 Fermion gas in the 
 harmonic trap. They are expressed in terms of the basic quantities $m$, $\omega _{\ell}$,
 $N$.
 One of them is clearly the Fermi energy $\epsilon _F$ according to (\ref{1.12}). At the Fermi 
 energy the filled Fermi sea has a spatial
 extension $2 L_F$ according to $ m \omega^2_l L_F^2/2=\hbar \omega _{\ell} (N-1/2)$ or
 
 \begin{equation}\label{2.1}
 L_F = \frac{1}{\alpha} \sqrt{2N-1} \equiv L_{n=N-1},
 \end{equation}
 
 a quantity frequently appearing later. The positions $z= \pm L_F $ are classical 
 turning points for a Fermion with energy $\epsilon _F$. The length $L_F$ is the
 largest length of the problem followed by $\alpha^{-1}$ which is associated with
 the zero point energy.
 
 Equation (\ref{A.10}) of the Appendix shows that a wave function $\psi _n$ 
 behaves as a standing wave with wave vector $k_n= \alpha \sqrt{2n+1}$ in the middle 
 of the trap provided $n \gg 1$. At the Fermi energy the wave number becomes 
 $k_F = \alpha \sqrt{2N-1}$. Together with (\ref{1.12}) this leads to $\epsilon _F 
 =\hbar ^2 k_F^2/2m$ as suggested in \cite{BR97} for the three--dimensional anisotropic
 case. The Fermi wave number $k_F^{-1}$ is the shortest length scale of the problem and 
 the Fermi energy the largest energy.
  
 What is the relation between $k_F$ and the particle density? 
 In a one--dimen\-si\-o\-nal Fermion gas of 
 spatial extension $2 L_F$ with periodic or open (infinite potential well) boundary conditions 
 the relation is in both cases 

 \begin{equation}\label{2.5}
 k_F^{(0)} = \pi n_0 = \pi \frac{N}{2 L_F}.
 \end{equation} 
 
In the present inhomogeneous situation $k_F$ increases as $N^{1/2}$ because the width 
$2L_F$ of the trap also increases as $N^{1/2}$. However,
we can discuss peak density $n_0^{(p)}$ and average density  $\bar{n}_0$ (or even higher 
moments of $n_0(z,N)$). The peak density is clearly found near  $z=0$. Using the asymptotic 
result (\ref{A.11})

\begin{equation}\label{A.11}
n_0 (z,N) = \frac{k_F}{\pi} + \frac{1}{2 \pi L_F} (1 - (-1)^N \cos 2 k_F z)
\end{equation}

gives

\begin{equation}\label{2.9}
n_0^{(p)} \ \sim \frac{k_F}{\pi}. 
\end{equation}

The sign $\sim$ denotes here and further on an asymptotic correspondence for $N \gg 1$.
Note that this asymptotic limit does not imply the semi--classical approximation.

From (\ref{2.9}) it is seen that the usual relation (\ref{2.5}) between $k_F$ and the 
one--dimensional particle density refers here to its peak value near the center. 

It is more difficult to discuss the average density $\bar{n}_0$ 
since an averaging length is needed.
Equations (\ref{2.7}) or (\ref{2.8}) only give the obvious sum rule

\begin{equation}\label{2.10}
\int^\infty _{- \infty}  dz\, n_0 (z, N)  = N.
\end{equation}

We thus resort to the semi--classical approximation where the local density is given 
\cite{BR97} by

\begin{equation}\label{2.13}
n_{sc} (z,N) \equiv \frac{k _F(z)}{\pi} = \frac{k_F}{\pi} \sqrt{1 - \left( \frac{z}{L_F} \right)^2}
\end{equation}

which is zero outside $|z| \leq L_F$. The corresponding average density clearly is
$\bar{n}_0  =k_F/4$ and is only slightly smaller than the peak density.

The sum rule for $n_{sc}$ gives $N -1/2$, i.e., half a Fermion is missing under the curve 
$n_{sc}$. This is due to the neglect of Fermion density leaking out of the classical 
region $|z| \leq L_F$ by tunneling. One might conclude that the number of Fermions in the 
oscillations is about one half. This is not correct for large Fermion number $N$ when the 
difference between $n_0(z,N)$ and $n_{sc}(z,N)$ near the boundaries becomes significant due 
to increasing oscillations in the exact density as detailed in the next Section. 

Finally, using formula (\ref{A.11}) immediately allows the identification of the ripples 
in $n_0 (z,N)$ near the center with the well known Friedel oscillations \cite{JF54} 
of wave number $2 k_F$ around an impurity in the degenerate Fermi sea.
In a naive interpretation these oscillations
result from the superposition of incoming and reflected parts of 
the uppermost wave function which both have a wave number $k_F$ near 
the center of the trap. A more subtle interpretation refers to the inherent
instability of the degenerate free Fermi gas towards static longitudinal 
perturbations of wave number $q=2k_F$. A well studied example are free electrons
(cf. e.g., \cite{PN66}). While in three dimensions only a logarithmic
singularity in the derivative with respect to $q$ appears in the susceptibility it 
becomes a
logarithmic singularity in one dimension due to perfect nesting. This causes 
charge and spin density instabilities when backscattering interactions are 
present. In bounded Luttinger liquids (cf. \cite {FG95,VYG00}) the interactions 
modify the divergence of the density oscillation near the boundary.

But even without interactions breaking of translational invariance by inhomogeneities 
like impurities and boundaries trigger density oscillations of wave vector $2k_F$. 
In one dimension this effect is well known for non--interacting Fermions with open 
boundary conditions. While in one
dimension the effect is most pronounced it is, nevertheless, possible 
to identify the oscillations in the isotropic density calculated in \cite{SW98} 
as three--dimensional Friedel oscillations. 

Since the one--dimensional Friedel-oscillations contain only about one atom it 
will be difficult to detect this effect experimentally. However, it is
conceivable to use an array of shorter micro traps each filled with a
reduced number of atoms.  The oscillations within each trap then add up
and lead to a total effect that is enhanced by the number of traps.
Using micro fabrication techniques it should be possible to combine 100
traps on one substrate leading to a signal that may become within reach of
advanced imaging techniques.

Friedel oscillations though difficult to observe are a fundamental 
property of the degenerate Fermi gas which eludes the semi--classical 
approximation.

\section{One--Particle Density Distribution}

With the help of (\ref{A.4}) the summation in (\ref{1.13}) can be performed for any $z$ and $N$ 
with the result:

\begin{equation}\label{2.7}
n_0 (z,N) = N \psi^2_N (z) - \sqrt{N (N+1)} \psi _{N+1} (z) \psi _{N-1} (z). 
\end{equation}   

Using the recurrence relations for the wave functions $\psi _n$ (cf. Appendix) this expression
can be brought into another useful form   

\begin{equation}\label{2.8}
 n_0 (z, N) = N \psi^2_{N-1} (z) - \sqrt{N (N-1)} \psi _N (z) \psi _{N-2}(z).
\end{equation}

This shows that the density distribution must be a polynomial of order $N-1$ in 
$\alpha^2 z^2$ times the exponential $\exp \{-\alpha^2 z^2 \}$ since the density is an even 
function of $z$.

The formulae (\ref{2.7}) and (\ref{2.8}) admit a number of exact 
conclusions as well as some remarkable asymptotic
results with respect to the Fermion number $N$.

Differentiating (\ref{2.7}) and (\ref{2.8}) with respect to $z$ and 
using the recurrence 
relations (\ref{A.1}) and (\ref{A.2}) gives

\begin{eqnarray}\label{3.1a}
\frac{\partial n_0 (z,N)}{\partial z} = - \alpha \sqrt{2 N}\,\psi _N(z)\, 
\psi _{N-1}(z), \\[4mm]\nonumber
\frac{\partial^2 n_0 (z,N)}{\partial z^2}
=2\alpha ^2 N\,[\psi^2 _N(z)-\psi^2 _{N-1}(z)].
\end{eqnarray}

This shows that the density distribution $n_0(z,N)$ has

\begin{itemize}
\item
$N$ maxima at the $N$ zeros $z_\nu^{(N)}$ of $ \psi _N(z)$,\newline 
($\nu=1,...,N$),
\item
$ N-1$ minima at the $N-1$ zeros $z_\nu ^{(N-1)}$ of $\psi _{N-1} (z)$, 
($\nu=1, ..., N-1 $).
\end{itemize}

As a consequence the minima of $n_0(z,N+1)$ touch the maxima of $n_0(z,N)$ at the points 
$z_\nu ^{(N)}$.
This is shown in Fig. 2 for $ N=5$. The area between $n_0 (z,6)$ and $n_0 (z,5)$ 
contains precisely one Fermion.
In this way the Pauli exclusion principle is optimally implemented.

The above considerations also show that about half a Fermion 
is contained in the ripples 
of the density distribution. The density at the maxima is given by

\begin{equation}\label{3.1}
n_0 (z_\nu ^{(N)},N) = N \,\psi ^2 _{N-1} (z_\nu ^{(N)}),
\end{equation}

and at the minima it is:

\begin{equation}\label{3.2}
n_0 (z_\nu ^{(N-1)},N) = (N-1) \,\psi ^2 _{N-2} (z_\nu ^{(N-1} ).
\end{equation}

Due to the knot theorems \cite {S67} the topological features inherent 
in the above statements carry over to arbitrary concave potentials.
Thus counting the number of maxima of the density distribution gives
the number of Fermions in any concave trap.

We now come to asymptotic results for $ N \gg 1 $. In practice, $N \approx 20$ is a 
good lower bound. In the asymptotic
region the powerful formula (\ref{A.5}) is available for the full range $|z| \leq L_F$. 
Inserting (\ref{A.5}) into (\ref{2.8}) gives

\begin{eqnarray}\label{3.3}
&n_0 (z,N) \sim k_F \left \{ \left( 1+ \frac{3}{4N} \right)\frac{(-t_{N-1})^{1/2}}{\sin 
\phi _{N-1}} \mbox{Ai}^2 (t_{N-1})
\right.\\ [4mm] \nonumber
&\left. -\left( 1 + \frac{1}{4N} \right) \, \frac{(t_N t_{N-2})^{1/4}}{\sqrt{ \sin \phi _N 
\sin \phi _{N-2}}} \,\mbox{Ai}(t_{N}) \mbox{Ai} (t_{N-2}) \right\}.
\end{eqnarray}

The functions $t_n (z) $ and $\phi _n (z)$ are defined in (\ref{A.6}) and (\ref{A.7}). 
The advantage of this formula lies in the fact that the indices of the wave functions moved
into the arguments of the Airy functions.

Evidently, the positions $z_\nu ^{(N)}$ of the maxima are now found from

\begin{equation}\label{3.4}
\mbox{Ai} \left( t_N(z_\nu ^{(N)})\right) = 0.
\end{equation}

We are interested in the positions of the last maximum, i.e., those lying in the neighbourhood 
of $L_F$. The asymptotic expansion of $t_N$ in the region $ z \leq L_F \sim (2N)^{1/2}$ is

\begin{equation}\label{3.5} 
-t_N \sim 2 \, N^{2/3} \left( 1 - \frac{z}{L_F} \right). 
\end{equation}

One also finds

\begin{equation}\label{3.6}
t_{(N-1)\pm 1} \sim t_{N-1} \mp N^{-1/3}.
\end{equation}

Provided $N^{1/3}$ is much larger than unity this leads to (the prime means the derivative)

\begin{eqnarray}\label{3.7}
n_0 (z_\nu ^{(N)}, N) \sim \alpha \sqrt{2}\, N^{5/6} \mbox{Ai}^2 
\left( t_{N-1} (z_\nu ^{(N)}) \right)\\[4mm]\nonumber
\sim \alpha \sqrt{2}\, N^{1/6} \mbox{Ai}^{'}(t_{N}(z_\nu^{(N)}))^2.
\end{eqnarray}

Specifically, we consider the last maximum at $z_N^{(N)}$. It corresponds to the first zero
of the Airy function $\mbox{Ai(t)}$ which is at $t_N (z_N ^{(N)})=-2,33...$\,.

This gives

\begin{equation}\label{3.8}
z_N ^{(N)} \sim L_F \left( 1- \frac{1.17}{N^{2/3}} \right) \equiv L_F - \Delta x_N.
\end{equation}

Note that
\begin{equation}\label{3.9}
\Delta x_N \sim \frac{1.17}{N^{2/3}} L_F \sim \frac{1.65}
{\alpha N^{1/6}},
\end{equation}

while the density at the maximum is

\begin{equation}\label{3.10}
n_0 (z_N^{(N)},N) \sim 0.7 \alpha \,N^{1/6}.
\end{equation}

In the same way the distance $l \equiv z_N^{(N)}-z_{N-1}^{(N)}$ between the 
last two maxima can be 
calculated. It determines the smallest local wave number $k^{(min)} \equiv
2 \pi/l$ which is found to be

\begin{equation}\label{3.10a}
k^{(min)} \sim 2 \pi \alpha \,\,0.8\, N^{1/6}.
\end{equation}

It needs more than $N=50$ Fermions to make $k^{(min)}$ less than half
the maximal wave number $k^{(max)} \equiv 2 k_F$ appropriate for the 
central part of the trap.

When we define a shrinking region $S(N)$ near $L_F$ according to

\begin{equation}\label{3.11}
\Delta x \equiv L_F-z = \frac{f}{\alpha N^{1/6}}
\end{equation}

with -- say -- $f$ varying from $1$ to $10$ then we have 

\begin{equation}\label{3.12}
-t_{N-1} \sim 2N^{2/3}  \frac{f}{\alpha L_F N^{1/6}} \sim \sqrt{2} \,f,
\end{equation}

which is independent of $N$.

In $S(N)$ the density (\ref{3.3}) can be drastically simplified to
read

\begin{eqnarray}\label{3.13}
n_0 (z,N) \sim \alpha \sqrt{2} \,N^{1/6} \left\{\mbox{Ai}^{'}(-t_{N-1}(z))^2 \right.
\\[4mm]\nonumber
\left.
-\mbox{Ai}(t_{N-1}(z))\mbox{Ai}^{''}(t_{N-1}(z)) \right\}.
\end{eqnarray}

Thus (\ref{3.7}) leads to a self similarity of the tails of 
the density for  
$z \in S(N)$ defined above: Calling $\tilde {n}_0(\Delta x, N)
= n_0(z=L_F-\Delta x,N)$ the graph of $\tilde{n}_0(\Delta x\in S(N_2),
N_2)$ maps 
precisely onto that of $ \tilde{n}_0(\Delta x \in S(N_1),N_1)$ when 
the density is 
rescaled according to $(N_1/N_2)^{1/6}$ and the position $\Delta x$
according to $(N_2/N_1)^{1/6}$. Note that $\tilde{n}_{sc}(\Delta x,N)$
also satisfies the scaling in $S(N)$.  

Finally we exploit the approximation (\ref{A.9}) which holds for large values of (-t), 
i.e., slightly away from the classical turning points $z=\pm L_F$. 
We are interested in results for $|z| < L_F$ which better the result (\ref{A.11}) valid in 
the very middle of the trap.

Using the expansion

\begin{equation}\label{3.16}
\phi _{(N-1)\pm 1} \sim \phi _{N-1} \pm \frac{1}{2N-1} \, \frac{z}{\sqrt{L_F^2 -z^2}}
\end{equation}

which requires distances $\Delta x/\alpha$ away from the boundary to be much larger than 
$N^{-1/6}$ we find  

\begin{eqnarray}\label{3.17}
&n_0 (z,N) = n_{sc} (z, N) +\frac{1}{2 \pi L_F \sqrt{1-(z/L_F)^2}}\\ [4mm]\nonumber
&-\frac{1}{2 \pi L_F}
\,\, \frac{\sin \left\{ (2N -1) \left(\sqrt{1- (z/L_F)^2}\,\,z/L_F 
- \arccos(z/L_F) \right) \right\} } 
{\sqrt{1-(z/L_F)^2}}.
\end{eqnarray}

Equation (\ref{3.17}) separates the slowly varying background $n_{sc}$ from an
increasing and spatially oscillating part due to quantum effects.  

It is also seen that the oscillating part increases towards the boundaries $z = \pm L_F$. 
A naive extrapolation would give an envelope

\begin{equation}\label{3.18}
E(z) \sim \frac{1}{\pi L_F \sqrt{1- (z/L_F)^2} } 
\end{equation}

of these oscillations which formally diverges near $L_F$ as  $(1- (z/L_F)^2)^{- 1/2}$. 
In view of the range of validity of (\ref{3.17}) this is, however, unwarranted. 
Nevertheless, 
it raises the question how the oscillating part diverges at the 
boundaries when $N$ diverges.

$N$ non--interacting Fermions in a one--dimensional box of width $L$ confined between  $z = 0$ 
and $z = L$, and with infinite barriers (open boundary conditions) have the density 
distribution (for $N \gg 1$)

\begin{equation}\label{3.19}
n_0(z,N) = \frac{k_F^{(0)}}{\pi} \left( 1 - 
\cot \left( \frac{\pi z}{L} \right) 
\frac{\sin 2 k_F^{(0)} z}{2 N} \right).
\end{equation}

Using (\ref{2.5}), i.e., $k_F^{(0)} = \pi N/L$
this gives an envelope for $z \ll L$ according to 

\begin{equation}\label{3.20}
E(z) \sim \frac{1}{\pi z} \equiv \frac{1}{\pi z^\delta}.
\end{equation}

We conjecture that in our case of a soft boundary the 
limiting behaviour near the right boundary for very large $N$ is

\begin{equation}\label{3.21}
E(z \leq L_F) \sim \frac{1}{2 \pi L_F(1 - z/L_F)^{\delta (N)}}.
\end{equation}

There is numerical evidence for  $\delta (N \rightarrow \infty) \rightarrow 1$, 
however, in a very slow approach ($\delta (N) \approx 1 - 1/\ln N$). This would imply
that the integrated absolute fluctuations

\[\delta N =\int^{LF}_{-LF}dz\,|n_0(z,N)-n_{sc}(z,N)|,\]

i.e., the number $\delta N$ of Fermions in the density oscillations
diverges logarithmically with $N$ as it does for ideal Fermions in a box.
The mathematical problem in clarifying this point lies in the enormous difficulty in 
subtracting out the oscillating part in  the boundary region 
$z \rightarrow L_F (N \gg 1)$ which is outside the
approximation (\ref{3.16}).

In case of interacting spinless one--dimensional Fermions in a box it is known that the
exponent $\delta$ is given by the coupling constant $K$ \cite {FG95}.

\section{Density Profile in Fourier Space}

With possible application to optical detection we discuss the 
Fourier transform of $n_0(z,N)$

\begin{equation}\label{4.1}
Fn_0(k,N) \equiv \int _{-\infty}^\infty dz\, e^{ikz} \,n_0(z,N).
\end{equation}

It can be evaluated exactly in the following way: The integral in  \cite{GR80} 
can be converted to the form

\begin{eqnarray}\label{4.3}
&\int _{-\infty}^\infty dz\,e^{ikz} \psi _m(z)\psi _n(z) 
\\[4mm]\nonumber 
&=e^{-\frac{k^2}{4 \alpha^2}}\,\,
(\frac{-k^2}{2 \alpha^2})^{(n-m)/2}\sqrt{\frac{m!}{n!}} \,L_m^{(n-m)}
(\frac{k^2}{2 \alpha ^2}),
\end{eqnarray} 

($n \ge m$) where $L_n$ denotes a Laguerre polynomial. Applying (\ref{4.3}) to
(\ref{2.8}) and using recursion relations for Laguerre polynomials \cite{AS70} 
gives

\begin{equation}\label{4.5}
Fn_0 (k,N)=e^{-\frac{k^2}{4\alpha^2}}\,L_{N-1}^{(1)}(\frac{k^2}{2 \alpha^2}). 
\end{equation}

The Fourier transform of the semi--classical expression (\ref{2.13}) 
can also be given in closed form involving a Bessel function:

\begin{equation}\label{4.6}
Fn_{sc}(k,N)=\frac{k_F}{k}\,\mbox{J}_1(\frac{k^2}{2 \alpha^2}). 
\end{equation}

Note the sum rules

\begin{equation}\label{4.7}
\int _{-\infty}^\infty dk\,Fn_0(k,N)=2k_F=\int _{-\infty}^\infty dk\,
Fn_{sc}(k,N), 
\end{equation}

and the limits

\begin{equation}\label{4.8}
Fn_0(k \rightarrow 0,N)=N,\quad Fn_{sc}(k \rightarrow 0,N)=N-\frac{1}{2}. 
\end{equation}

The basic difference between the exact result (\ref{4.5}) which takes care
of the ripples in the density profile and the semi--classical form (\ref{4.6})
is a hump somewhat below the wave number $2k_F$ as shown in Fig. 3.

\begin{figure}[h]
\begin{center}
\epsfig{figure=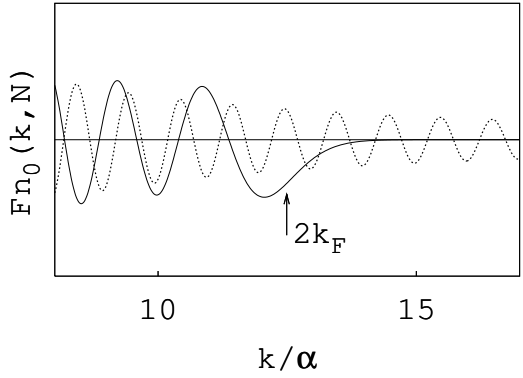,width=0.9\columnwidth} 
\end{center}
\caption{\small Part of the Fourier transformed density distribution function for $N=20$ 
non--interacting Fermions at zero temperature.
Note the small hump near the wave number $2 k_F$ due to Friedel oscillations which 
provide the shortest length scale of the problem. Dotted line is the semi--classical
approximation lacking that feature.}
\end{figure}

For larger wave numbers $Fn_0(k,N)$ drops to zero while $Fn_{sc}(k,N)$ shows a 
multitude of oscillations similar to that produced by a slit of spatial
width $2L_F$. For $k \ll 2 k_F$, however, the exact result and the semi--classical
approximation agree very well.

\section{One--Particle Momentum Distributions}

Even for a confined system one can define a momentum density distribution by

\begin{equation}\label{5.1}
p(k) \equiv \langle \hat{c}^+ _{k} \hat{c}_k \rangle.
\end{equation}

The operator $\hat{c}_k$ annihilates a Fermion with (continuous) momentum 
$\hbar k$. It can be decomposed into the fermionic annihilation operators 
for the harmonic oscillator according to

\begin{equation}\label{5.2}
\hat{c}_k = \sum _{n = 0}^\infty (-1)^n f_n^k \,\hat{c}_n,
\end{equation}

with the transformation function

\begin{equation}\label{5.3}
f_m^k = \frac{i^m}{\alpha} \psi _m ( z = \frac{k}{\alpha^2}).
\end{equation}

The momentum density $p_0$ of non--interacting Fermions in the harmonic trap and at zero
temperatures thus becomes

\begin{eqnarray}\label{5.4}
p_0 (k; N, T =0) &=& \sum _{m = 0}^\infty ( f_m^k )^* f_m^k \,\Theta (N-1-m)
\\[4mm\nonumber]
&=& \sum _{m = 0}^{N-1} (-1)^m  (f_m^k)^2,
\end{eqnarray}

leading to the remarkable result

\begin{eqnarray}\label{5.5}
p_0 (k; N, T =0) &=& \frac{1}{\alpha^2}\sum _{m = 0}^{N-1} \psi _m^2 (z = k/\alpha^2)
\\[4mm]\nonumber
&\equiv& \frac{1}{\alpha^2}\,n_0 ( z = k/\alpha^2; N, T =0 ). 
\end{eqnarray}

The momentum density is isomorphic to the particle density with $k_F$ replacing $L_F$.
Obviously it satisfies the general sum rule

\begin{equation}\label{5.6}
\int _{-\infty}^{\infty} dk\,p(k;N,T) = N.
\end{equation}

Alternatively, we can study the momentum probability

\begin{equation}\label{5.7}
P(k; N, T = 0) \equiv \int^\infty _ {- \infty} dz \,e^{ikz}\, \langle \hat{\psi}^+ (z) 
\hat{\psi}(0) \rangle ^{(N)}.
\end{equation}

which is also appropriate for symmetric confining potentials centered at $z=0$:

For non--interacting Fermions at zero temperature in the harmonic trap we can use
(\ref{A.4}) to find:

\begin{eqnarray}\label{5.8}
&&\langle \hat{\psi}^+ (z) \hat{\psi}(0) \rangle _0 ^{(N)} = \frac{\alpha}{\sqrt{\pi}}\, 
e^{- \alpha^2 z^2/2}\\[4mm]\nonumber
&&\sum _M\left\{ \delta _{N, 2M +1} L^{( \frac{1}{2}) }_M (\alpha^2 z^2) + \delta _{N, 2M} 
L^{( \frac{1}{2})}_{M-1}(\alpha^2 z^2) \right\}.
\end{eqnarray} 
 
In the limit $N \gg 1$ this reduces to the simple expression

\begin{equation}\label{5.9}
\langle \hat{\psi}^+ (z) \hat{\psi}(0) \rangle _0 ^{(N)} \sim 
\frac{\sin k_F z}{\pi z} + 0 \left( \frac{1}{\sqrt{N}} \right).
\end{equation}
 
The corresponding discrete momentum distribution for the harmonic oscillator 
is the well known step function 
 
\begin{equation}\label{5.10}
P_0 (k; N \gg 1, T =0) = \Theta (k_F - k)
\end{equation}

with $k = k_n = \alpha \sqrt{2 n + 1}$.

Some general remarks may be useful:
The centered momentum distribution (\ref{5.7}) can be expressed as 

\begin{equation}\label{5.11}
P(k; N,T) = \int^\infty _ {- \infty} dk' \,
\langle \hat{c}^+_k \hat{c}_{k'}\rangle. 
\end{equation}

In case of translational invariance the integrand in 
(\ref{5.11}) becomes  

\begin{equation}\label{5.12}
\langle \hat{c}^+_k \hat{c}_{k'}\rangle = P(k; N, T)\, \delta(k-k'). 
\end{equation}

Under periodic boundary conditions both definitions (\ref{5.1}) and
(\ref{5.7}) coincide.

\section{Free Expansion of Particle Density}

Expansion of a particle cloud is an important tool to investigate
Bose--Einstein condensates (c.f. \cite{HC96,KDSK99}). Detailed theories
are available for this expansion based on the Gross--Pitaevskii 
equation \cite{HC96,KSS96,CD96,HJC97,DMS97}. In the simplest case the
trap is suddenly switched off and the condensate expands freely.
Non--interacting Boson condensates display universal length 
scaling \cite{KSS96} in all spatial dimensions. The same has also been shown
recently for non--interacting Fermions \cite{BC00}.

When particle interactions dominate the kinetic energy the 
transverse scaling function in a highly anisotropic trap is 

\begin{equation}\label{7.1}
b_r(t)=\sqrt{1+\omega^2_r\, t^2},
\end{equation}

while the longitudinal expansion is more complicated \cite{CD96}.
This was confirmed experimentally in full detail in \cite{EMS98,ESS98}.

In accord with \cite{BC00} we find that a freely expanding degenerate one-dimensional gas of
non--interacting Fermions behaves according to (\ref{7.1})
with $\omega _{\ell}$ replacing $\omega _r$. 
The calculation is fully quantum mechanical and supplements the approach in 
\cite{BC00}. 

The quantity to be calculated is

\begin{equation}\label{7.2}
n(z,t) = \mbox{Tr}\,\hat{\rho}(t)\, \hat{\psi}^+(z)\,\hat{\psi} (z).
\end{equation}

where $\hat{\rho}(t)$ is the density operator of the freely expanding
gas. It is given in terms of the statistical operator $\hat{\rho}(0)$ 
immediately before the trap is opened at time $t=0$ by

\begin{equation}\label{7.3}
\hat{\rho} (t) = e^{-i\hat{H}_{00}t/\hbar}\hat{\rho}(0)\,
e^{i\hat{H}_{00}t/\hbar}. 
\end{equation}

The free expansion of non--interacting one--dimensional Fermions is governed 
by the Hamiltonian

\begin{equation}\label{7.5}
\hat{H} _{00} = \int^\infty _{-\infty} \, dk\, \frac{\hbar ^2 k^2}{2m}\,  
\hat{c} ^+ _k \hat{c} _k.
\end{equation}

The operators $\hat{c} ^+ _k$ and $ \hat{c}_k$ were introduced in conjunction
with (\ref{5.1}). Equation (\ref{7.2}) can also be written as

\begin{equation}\label{7.6}
n(z,t) = \mbox{Tr}\,  \hat{\rho} (0) \hat{\psi} ^+ (z,t) \hat{\psi} (z,t)
\end{equation}

with

\begin{eqnarray}\label{7.7}
\hat{\psi} (z,t) &=& e^{ i \hat{H}
_{00}t/\hbar}  \hat{\psi} (z) e^{-i\hat{H}_{00} t/\hbar}
\\[4mm]\nonumber
&=& \frac{1}{\sqrt{2 \pi}}
\int^\infty _{- \infty}\, dk \,e^{i (kz- \omega _k t)}\, \hat{c} _k
\end{eqnarray}

and

\begin{equation}\label{7.10}
\omega _k = \frac{\hbar k^2}{2m}.
\end{equation}

We now use (\ref{5.2}) and (\ref{5.3}) to find

\begin{eqnarray}\label{7.13}
n_0 (z,t) &=& \frac{1}{2 \pi} \int^\infty _{-\infty} \, dk \, dk '
e^{-iz (k-k') + i (\omega _k - \omega _{k'})t}\\ [4mm]\nonumber
&&\sum ^\infty _{m,n=0} (-1) ^{m+n} (f_m^k)^* \,f_n^{k'}\,
\mbox{Tr}\, \hat{\rho}(0)\hat{c} ^+_m \hat{c} _n.
\end{eqnarray}

For a harmonic trap which is initially in thermal equilibrium the statistical operator
$ \hat{\rho}(0)$ is (\ref{1.7}) and at zero temperature 

\begin{equation}\label{7.17}
\mbox{Tr}\, \hat{\rho}\,\hat{c} ^+ _m \hat{c} _n 
=\delta _{m,n} \,\Theta(N-1-m)
\end{equation}

holds. We thus find

\begin{eqnarray}\label{7.18}
&&n_0 (z,t;T=0) = \frac{1}{2 \pi \alpha^2}  
\int^ \infty _{-\infty}\, dk\,  dk' \\[4mm]\nonumber
&&e^{-iz(k-k')+i(\omega _k - \omega _{k'})t}\,\,
\left\{ \sum ^{N-1}_{m=0} \psi _m ( \frac{k}{\alpha^2})\, 
\psi _m (\frac{k'}{\alpha^2} ) \right \}.
\end{eqnarray}

The summation in curly brackets can be performed using (\ref{A.4}).
In order to proceed it is convenient to go over to the Fourier transform
and write out the oscillator eigenfunctions in terms of Hermite
polynomials. This leads to

\begin{eqnarray}\label{7.22}
&&Fn_0 (k_1,t;T=0) \equiv \int ^\infty _{-\infty}\, dz \, e^{ik_1z}\, n_0 (z,t; T=0)
\\[4mm] \nonumber
&&=\sqrt{ \frac{1}{\pi}}\,\,\frac{1}{2^N k_1 (N-1)!}\,e^{- \frac{k^2_1}{2\alpha ^2} 
(1-i\omega _{\ell} t)} \\[4mm]\nonumber  
&&\int^\infty _{-\infty} \, dk'\, e^{- \frac{k'^2}{\alpha^2} -\frac{k'k_1}{\alpha^2}
(1- i\omega _{\ell} t)}\,
\\[4mm]\nonumber
&&\left[ H_N ( \frac{k_1 +k'}{\alpha})\, \,H_{N-1} (\frac{k'}{\alpha})
 - (N \leftrightarrow (N-1)) \right].
\end{eqnarray}

Using \cite{GRb80} the integration can be performed giving

\begin{equation}\label{7.23}
Fn_0 (k,t;T=0) = e^{-\frac{k^2}{4 \alpha^2} (1+ \omega ^2 _{\ell} t^2)}\,
L^{(1)} _{N-1} ( \frac{k^2}{2\alpha^2} (1+ \omega ^2_{\ell} t^2)).
\end{equation}

This formula is isomorphic to (\ref{4.5}) with the inverse length
$\alpha $ being replaced by the rescaled value

\begin{equation}\label{7.25}
\alpha \rightarrow (1+\omega ^2_{\ell} t^2)^{-1/2}\alpha \equiv \alpha/ b(t).
\end{equation}

Since $Fn_0 (k,t)$ and $n_0(z,t)$ are related via a Fourier transformation
the final result is

\begin{eqnarray}\label{7.24}
n_0 (z,t;T=0) = \frac{1}{b(t)}\, n_0 ( \frac{z}{b(t)},N).
\end{eqnarray}

Thus free longitudinal expansion proceeds via a simple length rescaling 
involving the 
factor $b(t)$. In the course of time the initial density distribution Fig. 2
decreases and broadens according to the factor $b(t)$ but preserves its
topology including the Friedel oscillations which correspondingly increase 
their wave length.

In this picture it is assumed that the Fermions remain one--dimensional during
the expansion. If the transverse confining fields are also removed transverse 
expansion in any of the two equivalent transverse directions will also proceed  
according to (\ref{7.24}) taken for $N=1$
and with the scaling function (\ref{7.1}). This follows simply from the observation 
that each of the two ground state wave functions for the transverse directions 
correspond to a single one--dimensionally confined Fermion with $\omega _r$
in place of $\omega _{\ell}$.

\section{Summary}
We have calculated exactly one--particle properties of non--interacting
one--di\-men\-sio\-nal Fermions in a harmonic trap. These are the particle density
distribution including its free expansion when the trap is switched off
and also two momentum distribution functions. The exact calculability
can be traced back to two specific mathematical features of the eigenfunctions
of the harmonic oscillator namely that finite sums of bilinear expressions
can be performed and that Fourier transformation essentially reproduces 
an eigenfunction. Friedel oscillations in the particle density and its 
analogue in the momentum distribution as well as diverging density oscillations near the 
classical boundary are basic features of the degenerate 
one--dimensional ideal Fermi gas.\\
\\

{\bf Acknowledgements}: The authors thank Deutsche Forschungsgemeinschaft for financial
support.

\section{Appendix}

\newcounter{affix}
\setcounter{equation}{0}
\setcounter{affix}{1}
\renewcommand{\theequation}{\alph{affix}.\arabic{equation}}

The Appendix is a compilation of some mathematical formulae used in the derivation of the 
results given in the main part of the paper.
An important role is played by the recurrence relations for Hermite 
polynomials (cf. e.g., \cite{AS70}).
Here they are given as recurrence relations for the complete harmonic oscillator wave 
functions $\psi _n (z)$. These are  
 
\begin{equation}\label{A.1}
\sqrt{n+1} \,\psi _{n+1} (z) - \alpha z \sqrt{2} \,\psi _n (z) + 
\sqrt{n} \,\psi _{n-1} (z) = 0,
\end{equation}

\begin{equation}\label{A.2}
\frac{d}{dz} \psi _n (z) + \alpha^2 z \,\psi _n(z) - \alpha \sqrt{2n}\, \psi _{n-1} (z) = 0.
\end{equation}

The summation of the finite series (\ref{1.13}) is accomplished by means of

\begin{eqnarray}\label{A.4}
&&\sum^n_{m=0} \,\psi _m (z_1) \psi _m (z_2) \\[4mm]\nonumber
&&= \sqrt{\frac{n+1}{2}}\left[ \frac{\psi _{n+1} (z_1) \psi _n (z_2) - \psi _n (z_1) 
\psi _{n + 1} (z_2)}{\alpha (z_1 - z_2)} \right],
\end{eqnarray}

which is a conversion of a formula in \cite {BII}.
In (\ref{A.4}) the limit  $z_1 \rightarrow z_2$ can be 
performed and the resulting derivatives
converted into harmonic oscillator wave function using 
(\ref{A.2}) and (\ref{A.1}). This leads to (\ref{2.7}). An alternative derivation
applies induction to (\ref{2.8}) which is obviously true for $N=1$ utilizing 
the recurrence relation (\ref{A.1}).

A very useful asymptotic  ($n \gg 1$) expression for the wave 
functions can be extracted from \cite{AS70} (Chap. 19.7):

\begin{equation}\label{A.5}
\psi _n (z) \sim \sqrt{\alpha} \left(\frac{2}{n}\right)^{1/4} \left\{ \frac{(-t_n)^{1/4}}
{\sin^{1/2} \phi _n} \mbox{Ai} (t_n) \right\},
\end{equation}

with 

\begin{equation}\label{A.6}
- t_n = \left[ \frac{3}{2} \left( \frac{n}{2} + \frac{1}{4} \right) 
\left(2 \phi _n - \sin 2 \phi _n \right) \right]^{2/3}
\end{equation}

and 

\begin{equation}\label{A.7}
\cos \phi _n = \frac{z}{L_n}.
\end{equation}

$\mbox{Ai}$ is the Airy--function which oscillates for negative arguments. 
There is a continuation to positive arguments (the tunneling region)
which we will not discuss. 
 
The tilde $\sim$ denotes asymptotic expansion for large $n$ including 
all prefactors. Inside 
the trap, i.e., away from the classical borders, the form 

\begin{eqnarray}\label{A.9}
&&\psi _n (z) \sim \sqrt{\frac{\alpha}{\pi}} \left(\frac{2}{n}\right)^{1/4} \frac{1}{\sin^{1/2}
\phi _n}
\\[4mm]\nonumber 
&&\cos \left\{ \left( \frac{n}{2} + \frac{1}{4} \right) 
\left( \sin 2 \phi _n - 2 \phi _n \right) + \frac{\pi}{4}
\right\}
\end{eqnarray} 

is useful. It results by means of the asymptotic expansion of the 
Airy function $\mbox{Ai}(t)$ for 
$- t \gg 1$.
In the limit $|z| \ll L_n$ a further simplification occurs since $ \phi _n \longrightarrow 
\pi/2 -z/L_n$. This leads to (note: $n \gg 1$)

\begin{eqnarray}\label{A.10}
\psi _n \longrightarrow \left(\frac{2 \alpha^2}{n \pi^2}\right)^{1/4} \cos \left(k_n z - 
\frac{n \pi}{2} \right)
\end{eqnarray}

which is used in Sec. IV. The corresponding Fermion density well inside the trap is 
 
\begin{eqnarray}\label{A.11}
n_0 (z,N) = \frac{k_F}{\pi} + \frac{1}{2 \pi L_F} (1 - (-1)^N \cos 2 k_F z).
\end{eqnarray}

Here a small systematic error $1/(2 \pi L_F )$ of this approximation for $|z| \ll L_F $ 
has been subtracted to bring (\ref{A.11}) in line with the exact result.

\end{multicols}

\end{document}